\documentclass[preprint,letterpaper,12pt]{JHEP3}
\usepackage{graphics,epsfig}
\usepackage{axodraw}
\newcommand{\GeV}{\,{\rm GeV}}
\newcommand{\MeV}{\,{\rm MeV}}

\newcommand\lsim{\mathrel{\rlap{\lower4pt\hbox{\hskip1pt$\sim$}}
    \raise1pt\hbox{$<$}}}
\newcommand\gsim{\mathrel{\rlap{\lower4pt\hbox{\hskip1pt$\sim$}}
    \raise1pt\hbox{$>$}}}
\def\bea{\begin{eqnarray}}
\def\eea{\end{eqnarray}}
\def\ba{\begin{array}}
\def\ea{\end{array}}
\def\bec{\begin{center}}
\def\ec{\end{center}}
\def\nn{\nonumber}

\def\la{\langle}
\def\ra{\rangle}

\def\64{\rm SO(6) \times SO(4)}

\def\f{\frac}
\def\a{\alpha}
\def\b{\beta}

\preprint{OHSTPY-HEP-T-04-002}
\title{\huge Quark Mass Textures and $\sin 2 \beta$}
\author{Hyung Do Kim$^{a,b}$, Stuart Raby$^a$ and Leslie Schradin$^a$\\
$^a$Department of Physics, The Ohio State University,\\
174 W. 18th Ave., Columbus, Ohio 43210, USA\\
$^b$School of Physics, Seoul National University,\\
Seoul, 151-747, Korea\\
\\ E-mail: \email{hdkim,raby,schradin@mps.ohio-state.edu}}

\abstract{Recent precise measurements of $\sin 2 \beta$ from the B-factories (BABAR and BELLE) and a better known
strange quark mass from lattice QCD make precision tests of predictive texture models possible.
The models tested include those hierarchical N-zero textures classified by Ramond, Roberts and Ross, as well as
any other hierarchical matrix Ansatz with non-zero $12 = 21$ and vanishing 11 and 13 elements. We calculate the
maximally allowed value for $\sin 2 \beta$ in these models and show that all the aforementioned models with
vanishing 11 and 13 elements are ruled out at the 3$\sigma$ level. While at present $\sin 2\b$ and $|V_{ub}/V_{c
b}|$ are equally good for testing N-zero texture models, in the near future the former will surpass the latter in
constraining power. }


\begin{document}

\section{Introduction}

The problem of understanding the origin of fermion masses has persisted for more than twenty years
\cite{Weinberg:hb,Wilczek:uh,Fritzsch:1977za}.  Models predicting relations between fermion masses and mixing
angles may give insights into possible solutions to this problem.  On the other hand, testing theories of fermion
masses requires precision data; the accuracy of which has been severely limited by theoretical uncertainties
inherent in QCD. In particular, light quark masses and some CKM elements have been difficult to measure with
precision.  Recent results from B-factories combined with advances in the theory of heavy quarks as well as in
lattice QCD have reduced the errors considerably for these observables~\cite{Battaglia:2003in,Hocker:2001xe}.
Current measurements of $|V_{us}|$ and $|V_{cb}|$ have errors at the 2\% level~\cite{Battaglia:2003in}.  $\sin
2\b$ is now known to 6.5\% from experiments on the asymmetry in B decays~\cite{Hocker:2001xe}. Even $\left|
V_{ub}/V_{cb} \right|$, whose errors largely come from non-perturbative QCD effects, has now been determined to
about 10\%~\cite{Battaglia:2003in}. In the mass sector, the most important improvement has been in $m_s$, whose
uncertainty has decreased from 50\% to 12\% over the last ten years.  Moreover, lattice QCD results with light
dynamical quarks indicate that the strange quark mass is much lighter than previously
thought~\cite{Gupta:2001cu}.

In a pioneering work Hall and Rasin~\cite{Hall:1993ni} showed that the relation \bea |V_{ub}/V_{cb}| = &
\sqrt{m_u/m_c} & \eea is obtained for any hierarchical texture with vanishing $11,\ 13,\ 31$ elements.    Roberts
et al.~\cite{Roberts:2001zy} then re-analyzed these textures, using more recent data, and concluded that such
textures were disfavored, i.e. disagreeing with data at about 1$\sigma$ (see also ~\cite{Fritzsch:2001nv}).
Whereas the addition of a small $13 \approx 31$ element gave good fits to the data. They also studied
non-hierarchical asymmetric textures, with vanishing $13, \; 31$ elements but satisfying the ``lopsided" relation
$32 \sim 33$.  Good fits to the data were also obtained in this case. The strongest constraint, in their
analysis, came from the observable $|V_{ub}/V_{cb}|$, which at the time had an uncertainty
${\cal O}$(22\%).\footnote{Although Roberts et al. assumed an uncertainty half this size~\cite{Roberts:2001zy}.}  On the
other hand, the value of $\sin 2\b$ was not well known and, in fact, the central value was much lower than it is
now. However, with symmetric textures with non-zero $13 = 31$ elements, they predicted $\sin 2\b$ to be near its
present experimental value.

With the significant improvement in the data,  we feel that it is once again a good time to re-analyze quark mass
textures.  In this paper we study hierarchical textures 
satisfying $12 = 21$. We
find that such textures, with vanishing $11, \ 13, \ 31$ elements, are now excluded by 3$\sigma$. In the present
study, constraints from both $|V_{u b}/V_{c b}|$ and $\sin2\beta$ are equally strong. However, in the near future
$\sin2\beta$ may provide the most stringent constraint.  For example, it is expected that the experimental
precision of $\sin2\beta$ will greatly improve (by a factor of 2), perhaps by 2006 when an expected 500 fb$^{-1}$
of data will have been tabulated by both BaBar and Belle~\cite{Raven:2003gs}. On the other hand, the experimental
uncertainties associated with $|V_{u b}/V_{c b}|$ (limited by uncertainties in $|V_{u b}|$) may require a Super-B
factory (perhaps by 2010~\cite{Burchat:2001de}) to obtain a similar factor of 2 reduction~\cite{Ligeti:2003hp}.

In Section \ref{sec:data} we tabulate the latest data on quark masses and mixing angles.  Then in Section
\ref{sec:setup} we present the relevant approximations used when diagonalizing hierarchical fermion mass
textures.   In Section \ref{sec:2x2}, as a warm-up, we consider the oldest successful texture describing the
lightest two quark families.   We then obtain our main result in Section \ref{sec:13=0} where we study
hierarchical 3 $\times$ 3 quark mass textures satisfying  $12 = 21$ with $11 = 13 = 31 \equiv 0$. Finally in
Section \ref{sec:13neq0} we show that good fits to the data can be obtained with non zero $13, \ 31$ elements. In
particular we focus on two 5-zero texture models considered by Ramond et al.~\cite{Ramond:1993kv}.

\section{Data \label{sec:data}}

In this Section we tabulate the present data for CKM elements and quark masses used in our analysis.

\subsection{CKM elements}

We take the CKM element values from \cite{Battaglia:2003in} and $\sin 2\b$ from \cite{Hocker:2001xe}.

\bea
\left| V_{us} \right| & = & 0.2240 \pm 0.0036 \\
\left| V_{cb} \right| & = & (41.5 \pm 0.8) \times 10^{-3} \nn \\
\left| V_{ub} \right| & = & (35.7 \pm 3.1) \times 10^{-4} \nn \\
\left| \f{V_{ub}}{V_{cb}} \right| & = & 0.086 \pm 0.008 \nn \\
\sin 2 \b & = & 0.739 \pm 0.048 \nn \eea The errors on $\sin 2\b$ are mostly statistical while those on $\left|
\f{V_{ub}}{V_{cb}} \right|$ are largely theoretical.  As more data is taken, $\sin 2\b$ will become more
precisely known, but the precision in $\left| \f{V_{ub}}{V_{cb}} \right|$ will likely remain at the 10\% level
for some time.

\subsection{Masses}

Due to strong interactions, the masses of the light quarks are not known well. The most precise estimates of
$m_u/m_d$ and $m_s/m_d$ come from chiral perturbation theory.  There is some disagreement in the literature on
the sizes of the errors of the light quark mass ratios~\cite{Hagiwara:fs}.  For this reason, we take $m_u/m_d$
and $m_s/m_d$ from~\cite{Leutwyler:1996qg} with doubled errors.

Recent lattice QCD calculations with dynamical quarks have improved our knowledge of $m_s$, previously the least
known of the light quarks. We use the unquenched lattice QCD result with $n_f =2$ for $m_s$~\cite{Gupta:2001cu}
and double the error to account for the discrepancy with the sum rule result. The central value (Eqn.
\ref{eq:masses}) is near the low end of the range given by the PDG~\cite{Hagiwara:fs}.  The preliminary result
with $n_f = 2 + 1$ indicates that the strange quark might be even lighter.

For the charm quark mass we use a quenched lattice QCD result since an unquenched calculation is not yet
available. Quenching errors are known to be about 25\% for the strange quark mass and 1 to 2\% for the bottom
quark mass. Because of the mass hierarchy, it is expected that the quenching error on $m_c$ will lie somewhere
between these two bounds. Thus we take the lattice QCD result with a (probably conservative) 10 percent
systematic (quenching) error as in~\cite{Battaglia:2003in} and double it.

We use the bottom quark mass from~\cite{Bauer:2002sh} and the top quark pole mass from the Particle Data Group
(PDG) 2003~\cite{Hagiwara:fs}.
\bea \f{m_u}{m_d} & = & 0.553 \pm 0.043 \times 2 \label{eq:masses}  \\
\f{m_s}{m_d} & = & 18.9 \pm 0.8 \times 2 \nn \\
m_s (2\GeV) & = & 89 \pm 11 \times 2 \MeV \nn \\
m_c (m_c) & = & 1.30 \pm 0.15 \times 2 \GeV \nn \\
m_b (m_b) & = & 4.22 \pm 0.09 \GeV. \nn \\
M_t ({\rm pole}) & = & 174.3 \pm 5.1 \GeV \nn \\
m_t(m_t) & = & 165 \pm 5 \GeV \nn \eea
All running mass parameters are defined in the $\overline{\rm MS}$ scheme.
We note here that the doubled errors we are using almost incorporate the bounds found in the PDG
\cite{Hagiwara:fs}.

For the purposes of further analysis we compare to fermion masses evaluated at $M_Z$.  We define the
renormalization factor
\bea
\eta_i \equiv \left\{
\ba{cl}
\f{m_i(M_Z)}{m_i(m_i)} & \; {\rm for} \;\; i = c,\ b,\ t \\
& \\
\f{m_i(M_Z)}{m_i(2 \ {\rm GeV})} & \; {\rm for} \;\; i = u,\ d,\ s.
\ea
\right.
\eea
At two loops in QCD we find
\bea
\eta_c = 0.56, \; & \eta_b = 0.69, \; & \eta_t = 1.06 \\
\eta_u \; = \; \eta_d \; & = \; \eta_s \; = \; 0.65 . & \nn
\eea

\section{Setup \label{sec:setup}}

This section introduces our notation for the Yukawa and CKM matrices. Quark masses are expressed in terms of Weyl
spinors as \bea -{\cal L} & = & Q_i Y^U_{ij} \la H_u \ra u^c_j + Q_i Y^D_{ij} \la H_d \ra d^c_j, \eea where
$i,j=1,2,3$ are the family indices, $Q=\left(\ba{c} u \\ d \ea \right)$ is the left-handed quark doublet, $u^c$
and $d^c$ are the left-handed anti-up and -down quarks, and $H_{u}$ and $H_{d}$ are the up and down Higgs fields.
To keep track of phases in $Y^U$ and $Y^D$, which in general are complex, we define $\phi^U_{ij} \equiv \arg
Y^U_{ij}$ and $\phi^D_{ij} \equiv \arg Y^D_{ij}$. Two unitary matrices $V_U$ and $U_{U^c}$ diagonalize the up
quark Yukawa matrix by \bea V_U Y^U U_{U^c} & = & Y^U_{\rm Diag}. \nn \eea Similarly, $V_D$ and $U_{D^c}$
diagonalize $Y^D$. We assume that the mass matrices are hierarchical \cite{Hall:1993ni} which means \bea
\left|\f{Y^{(U,D)}_{(23,32)}}{Y^{(U,D)}_{33}}\right| & \ll & 1, \ \ \ \
\left|\f{{\tilde Y}^{(U,D)}_{22}}{Y^{(U,D)}_{33}}\right| \ll 1, \\[2mm]
\left|\f{Y^{(U,D)}_{(12,21)}}{{\tilde Y}^{(U,D)}_{22}}\right| & \ll & 1, \ \ \ \
\left|\f{Y^{(U,D)}_{11}}{\tilde{Y}^{(U,D)}_{22}}\right| \ll 1,  \nn \\[2mm]
\left|\f{Y^{(U,D)}_{13} \ Y^{(U,D)}_{31}}{Y^{(U,D)}_{33}}\right|& \ll & \left|\f{Y^{(U,D)}_{12} \
Y^{(U,D)}_{21}}{\tilde{Y}^{(U,D)}_{22}}\right|, \nn \eea 
where \bea \tilde{Y}_{22} & \equiv & Y_{22} - \f{Y_{23} \
Y_{32}}{Y_{33}} \equiv \left| \tilde{Y}_{22} \right| \ e^{i \tilde \phi_{2 2}} \eea is used since we will
consider both of the following cases: \bea 0 = |Y_{22}| < \left| \f{Y_{23} \ Y_{32}}{Y_{33}} \right| \;\;\; {\rm
and} \;\;\; 0 \neq |Y_{22}| \gtrsim \left| \f{Y_{23} \ Y_{32}}{Y_{33}} \right|. \eea  Note, an approximately
symmetric matrix ansatz will be hierarchical unless the known CKM mixing angles come as a surprising cancellation
of two large angles for up and down quarks. We also note that there are ``lopsided" textures in the literature
with $|Y_{32}| \sim |Y_{33}|$ which do not fulfill the conditions described above. We do not consider such
asymmetric textures in this paper. For an analysis of these models, see \cite{Roberts:2001zy}.

With no order one off-diagonal terms in $Y^U$ or $Y^D$, we can neglect higher powers of the mixing angles and
keep $\cos \theta \simeq 1$. This makes the mixing matrices quite simple.
\bea V_U & = & \left( \ba{ccc} 1 & -s^U_{12} & 0 \\
s^{U*}_{12} & 1 & 0 \\
0 & 0 & 1 \ea \right)
\left( \ba{ccc} 1 & 0 & -s^U_{13} \\
0 & 1 & 0 \\
s^{U*}_{13} & 0 & 1 \ea \right)
\left( \ba{ccc} 1 & 0 & 0 \\
0 & 1 & -s^U_{23} \\
0 & s^{U*}_{23} & 1 \ea \right) .\eea   $V_D$ has the same form.  The $s^U_{ij}$ can be considered as generalized
mixing angles carrying both a real mixing angle and a phase.

The CKM matrix $V_{CKM} = V_U^* \, V_D^T$ is then given by \bea V_{CKM} & = & \left( \ba{ccc} 1 & s^*_{12} +
s^{U*}_{13} s_{23} & - s^{U*}_{12} s^*_{23} + s^*_{13} \\
-s_{12} - s^D_{13} s^*_{23} & 1 & s^*_{23} + s^U_{12} s^*_{13} \\
 s^D_{12} s_{23} -s_{13} & -s_{23} - s^{D*}_{12} s_{13} &
1 \ea \right), \eea where \bea s^U_{12} & \simeq &
\f{Y^U_{12}}{\tilde{Y}^U_{22}}, \ \ \ \
 s^D_{12} \simeq \f{Y^D_{12}}{\tilde{Y}^D_{22}}, \\
s^U_{13} & \simeq & \f{Y^U_{13}}{Y^U_{33}}, \ \ \ \ s^D_{13}
\simeq \f{Y^D_{13}}{Y^D_{33}},
\nn \\
s^U_{23} & \simeq & \f{Y^U_{23}}{Y^U_{33}}, \ \ \ \ s^D_{23}
\simeq \f{Y^D_{23}}{Y^D_{33}},
\nn \\
s_{23} & \equiv & s^D_{23} - s^U_{23}, \ \ \ \ s_{12} \equiv s^D_{12} -
s^U_{12}, \ \ \ \ s_{13} \equiv s^D_{13} - s^U_{13}. \nn \eea

The above approximations work well, {\em except} in the case of first and second generation mixing. For
$|Y^D_{11}|=0$, $|Y^D_{12}| = |Y^D_{21}|$, which we will address in this paper, $|s^D_{12}| = \left|
\frac{Y^D_{12}}{\tilde{Y}^D_{22}} \right| \simeq \sqrt{\f{m_d}{m_s}} \left( 1 + {\cal O} \left( \f{m_d}{m_s}
\right)\right)$. To capture the rather large ${\cal O} \left( \f{m_d}{m_s} \right) \sim 5\%$ correction properly,
we must include at least through ${\cal O} (|s^D_{12}|^2)$ in $c^D_{12} \equiv \sqrt{1 - |s^D_{12}|^2}$. We need
do this only for the first and second generation mixing in the down sector, for all other mixing angles are small
enough that the errors from our approximations are below 1\%. (e.g., $\f{m_u}{m_c} \sim 2 \times 10^{-3}$).

An exact diagonalization of the 2 $\times$ 2 submatrix for down quarks gives \bea |s^D_{12}| & = &
\sqrt{\f{m_d}{m_s + m_d}}  \eea and thus $|s_{12}|$ ($ \subset |V_{us}|$) is given by,\footnote{In Eqn.
\ref{eq:s12} the second factor includes the cosine of the down quark mixing angle, i.e. $s^U_{1 2} \rightarrow
s^U_{1 2} \; c^D_{1 2}$. \label{fn:cosine}} \bea |s_{12}| & = & \sqrt{\f{m_s}{m_s+m_d}} \left|
\sqrt{\f{m_d}{m_s}}- e^{i\phi} \sqrt{\f{m_u}{m_c}} ) \right|  \label{eq:s12} \eea  where $\phi \equiv
(\phi_{12}^U - \tilde \phi_{22}^U) - (\phi_{12}^D - \tilde \phi_{22}^D)$.  For $ \ Y^U_{12} = Y^U_{21}=0$ we have
\bea |s_{12}| & = & \sqrt{\f{m_d}{m_s+m_d}} . \eea Furthermore, $|s_{23}| \sim 0.04 \ll |s_{12}|$ and we have
$V_{us} = s^*_{12} + s^{U*}_{13} s_{23} \simeq s^*_{12}$ within a 1\% error. Therefore, $V_{us} = s^*_{12}$ is
used.

We will also need $\b$, the most precisely measured angle within the unitarity triangle. In terms of CKM matrix
elements: \bea \b & = & {\rm arg} \left( -\f{V_{cd} V^*_{cb}}{V_{td} V^*_{tb}} \right). \eea

\section{Analysis}

In this section we confront the data.  We first consider the 2 $\times$ 2 subsector of light quarks and show that
good fits to the data are obtained.  In section \ref{sec:3x3} we extend the analysis to the full 3 $\times$ 3
case.

\subsection{2 $\times$ 2 light quark matrices \label{sec:2x2}}

The Cabbibo angle, in the original texture by Weinberg \cite{Weinberg:hb}, is generated solely by the down quark
Yukawa matrix.  The Yukawa matrices are given by \bea Y^U = \left( \begin{array}{cc} A & 0 \\ 0 & B \end{array}
\right) \;\;\;\; Y^D = \left(
\begin{array}{cc} 0 & C \\ C & D \end{array} \right) \eea where the parameters $A,\ B,\ C,\ D$ can be taken to be
real without loss of generality. Hence \bea |V_{us}| & = & \sqrt{\f{m_d}{m_s+m_d}} \label{eq:Vus1} \eea which
works surprisingly well since the observed $|V_{us}| = 0.2240 \pm 0.0036$ is just the same as
$\sqrt{\f{m_d}{m_s+m_d}} = 0.224 \pm 0.004$.

On the other hand, if $Y^U$ is taken to have the same form as $Y^D$ then a non-removable phase enters in the
determination of $|V_{us}|$.  We then find \bea |V_{us}|& = & \sqrt{\f{m_s}{m_s + m_d}} \left|\sqrt{\f{m_d}{m_s}}
- e^{i\phi} \sqrt{\f{m_u}{m_c}}\right|. \label{eq:Vus2} \eea  
Note, using the central value of the quark masses,
$V_{us}$ has the right value for $\phi \sim \f{\pi}{2}$.  
However, due to the large uncertainties in the light
quark masses, the value of $\phi$ is not significantly constrained.

In what follows, we always assume $|Y^D_{12}| = |Y^D_{21}| \neq 0$ and $Y^{(U,D)}_{11} = 0$.
These conditions imply that $|V_{us}|$ is given by either equation (\ref{eq:Vus1}) or (\ref{eq:Vus2}).
We will be using these expressions for $|V_{us}|$ throughout the rest of the analysis.

\subsection{3 $\times$ 3 quark matrices \label{sec:3x3}}

Over the next two subsections we address two categories of 3 generation quark textures. In section
\ref{sec:13=0}, we consider textures with zero 11 and 13 elements and symmetric 12 and 21 elements. We show that
all such textures, provided they are hierarchical, are ruled out. These include the 5-zero models I, II, and
IV, nominally consistent with previous data, classified by Ramond et. al. in~\cite{Ramond:1993kv} and listed in
Table \ref{t:5textures}.\footnote{6-zero symmetric texture models are already ruled out~\cite{Ramond:1993kv}.}
Section \ref{sec:13neq0} covers textures with non-zero 13 elements. We choose to study models III and V of Table
\ref{t:5textures}, the most constrained of these models, to illustrate that texture models of this type with 5 or
fewer zeros are consistent with the data.

\begin{table}
\caption[8]{ The symmetric 5-zero textures classified in \cite{Ramond:1993kv}.
The number of zeros (here, 5) refers to the total
number of zero elements in the upper-right and diagonal portions of the
up and down matrices.}
\label{t:5textures}
$$
\begin{array}{lcc} &  Y^U &  Y^D  \\
\hline
I & \left( \ba{ccc} 0 & X & 0 \\
X & X & 0 \\
0 & 0 & X \ea \right) &
\left( \ba{ccc} 0 & X & 0 \\
X & X & X \\
0 & X & X \ea \right) \\
\hline
II & \left( \ba{ccc} 0 & X & 0 \\
X & 0 & X \\
0 & X & X \ea \right) &
\left( \ba{ccc} 0 & X & 0 \\
X & X & X \\
0 & X & X \ea \right) \\
\hline
III & \left( \ba{ccc} 0 & 0 & X \\
0 & X & 0 \\
X & 0 & X \ea \right) &
\left( \ba{ccc} 0 & X & 0 \\
X & X & X \\
0 & X & X \ea \right) \\
\hline
IV & \left( \ba{ccc} 0 & X & 0 \\
X & X & X \\
0 & X & X \ea \right) &
\left( \ba{ccc} 0 & X & 0 \\
X & X & 0 \\
0 & 0 & X \ea \right) \\
\hline
V & \left( \ba{ccc} 0 & 0 & X \\
0 & X & X \\
X & X & X \ea \right) &
\left( \ba{ccc} 0 & X & 0 \\
X & X & 0 \\
0 & 0 & X \ea \right) \\
\hline
\end{array}
$$
\end{table}

\subsubsection{Models with $11=13=0$ and $12=21\neq0$ \label{sec:13=0}}

We consider here hierarchical texture models with $Y^{(U,D)}_{11} = Y^{(U,D)}_{13} = 0$, $|Y^U_{12}| = |Y^U_{21}|
\neq 0$, and $|Y^D_{12}| = |Y^D_{21}| \neq 0$ which include type I, II and IV of \cite{Ramond:1993kv}. By the
method presented in the Appendix, we see that there are 2 non-removable phases in the mass matrices for I and IV
and 3 phases are non-removable 
for II.\footnote{Note, in our analysis we specifically use the phases from I, II
and IV 5-zero models.  Nevertheless, since we are interested in $\left| V_{u b}/V_{c b} \right|$ and in the
maximum value for $\sin2\beta$ our results hold for all models with $13 = 31 = 0$ and $12 = 21 \neq 0$.} The
location of these phases can be chosen as follows: \bea
\begin{array}{ccc}
{\rm I} \;\;\;\;\; & \; \phi^D_{22} & \; \phi^U_{22} \\
{\rm II} \;\;\;\;\; & \; \phi^D_{22} & \;\;\; \phi^U_{23} \;\;\;\;\; \phi^U_{32} \\
{\rm IV} \;\;\;\;\; & \; \phi^D_{22} & \; \phi^U_{32}
\end{array} \nn
\eea

\begin{figure}[t]
\begin{center}
\begin{picture}(100,200)
\put(10,105){\makebox(50,50){\includegraphics[height= 3.5
in]{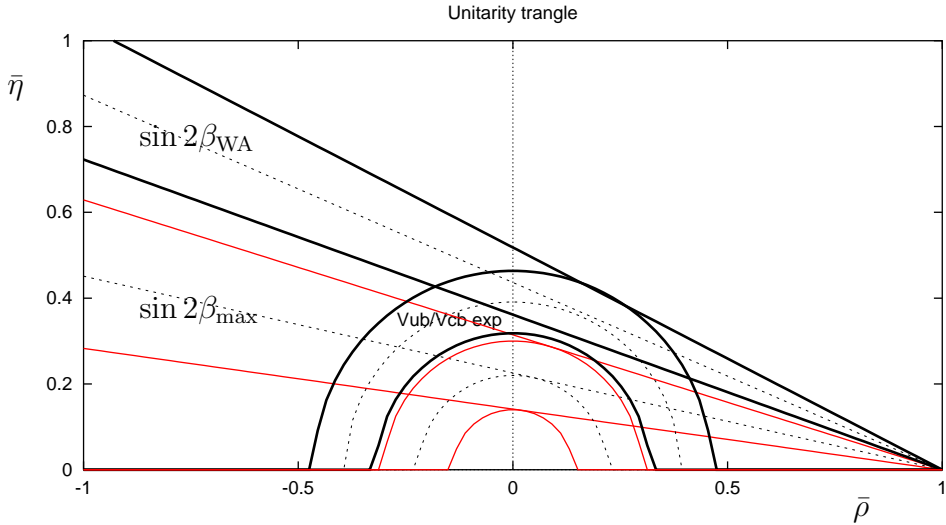}}} \put(-150,160){$\bar\eta$} \put(-100,140){$\sin
2\b_{\rm WA}$} \put(-100,75){$\sin 2\b_{\rm max}$}
\put(170,0){$\bar\rho$}
\end{picture}
\caption{Comparisons between experiment and predictions of hierarchical
$11=13=0$ and $12=21$ models.
Shown are the 2$\sigma$ lines for the predicted $\sin 2\b_{\rm max}$
(in red or grey) and the measured $\sin 2\b$ (in black).
The red (or grey) arcs correspond to the 2$\sigma$ range of the
predicted $|V_{ub}/V_{cb}|$, while the black arcs show the 2$\sigma$
range of the experimental value. The dotted lines and arcs
are the corresponding central values. There is no overlap between
the predictions and experiment for either observable.}
\label{fig:sin2bI}
\end{center}
\end{figure}

$Y^U_{13} = Y^D_{13} = 0$ implies $s_{13} =0$, and the CKM matrix becomes extremely simple: \bea V_{CKM} & = &
\left( \ba{ccc} 1 \;\;\; & s^*_{12} \;\; &  - s^{U*}_{12} \ s^*_{23} \\
-s_{12} \;\;\; & 1 \;\; & s^*_{23} \\
s^D_{12} \ s_{23} \;\;\; & -s_{23} \;\; & 1 \ea \right)  \eea
\bea s^U_{12} & = &  \f{Y^U_{12}}{\tilde Y^U_{22}}
\simeq \sqrt{\f{m_u}{m_c}} e^{-i \tilde \phi^U_{22}} \\[2mm]
s^D_{12} & \simeq &
 \sqrt{\f{m_d}{m_s + m_d}} e^{-i \tilde \phi^D_{22}} \nn \\[2mm]
\tilde \phi_{22} & \equiv & \tilde \phi^D_{22} - \tilde \phi^U_{22} . \nn \eea We now use these angles to find the CKM
elements and $\b$ in terms of mass ratios and phases. In the case of $|V_{us}|$, we use the expression found in
equation (\ref{eq:Vus2}) (also see footnote $^2$ with regards to $\b$).  \bea |V_{us}| & = &  \sqrt{\f{m_s}{m_s +
m_d}}
\left|\sqrt{\f{m_d}{m_s}} - e^{-i \tilde \phi_{22}} \sqrt{\f{m_u}{m_c}}\right| \label{eq:Vus} \\[2mm] 
\left|\f{V_{ub}}{V_{cb}}\right| & = & |s^U_{12}|
\simeq \sqrt{\f{m_u}{m_c}} \nn \\[2mm]
\b & = & \arg \left( \f{s_{12}}{s^D_{12}} \right) = \arg \left( 1- \f{s^U_{12} \ c^D_{1 2}}{s^D_{12}} \right)
\simeq \arg \left( 1- r e^{i \tilde \phi_{22}} \right) \nn \eea where \bea r \equiv
\sqrt{\f{m_u}{m_c}\f{m_s}{m_d}}. \eea  Since $r \sim 1/4 \ll 1$, $\b$ is restricted to be a small angle. The
maximum value for $\b$, i.e.  $\b_{\rm max}$, is given by
\bea \sin \b_{\rm max} & = & r, \\
\sin 2 \b_{\rm max} & = & 2 r \sqrt{ 1-r^2 } . \nn  \eea  Using the data for the quark masses, we get $r = 0.22
\pm 0.04$. Thus this texture model results in  \bea \sin 2 \b_{\rm max} & = & 0.43 \pm 0.08 . \eea  Given the
world average $\sin 2 \b_{\rm WA} = 0.739 \pm 0.048$, 
this texture is ruled out by at least 3.4$\sigma$.{\footnote{We add errors in quadrature and compare it to the difference between the central values $|\sin
2\b_{\rm WA} - \sin 2 \b_{\rm max}|$.}}  Using equation (\ref{eq:Vus}), we also find that this model predicts
\bea \left| \f{V_{ub}}{V_{cb}} \right| = 0.051 \pm 0.009. \eea Comparison with the measured value, $0.086 \pm
0.008$, leads to a difference of 2.9$\sigma$.   In Figure \ref{fig:sin2bI} we show the discrepancy between the
experimental data and the predictions of $\sin 2\b_{\rm max}$ and $|V_{ub}/V_{cb}|$ for these textures. The
2$\sigma$ error bars for the experimental and predicted values do not overlap for either observable.  Note, the
more precise $\sin 2\b$ measurement is slightly more effective in constraining these models than
$|V_{ub}/V_{cb}|$.

We have shown in this section that even with a conservative doubling of the light quark masses listed in equation
(\ref{eq:masses}), all hierarchical texture models with 11 = 13 = 0 and 12 = 21 are ruled out at the 3$\sigma$
level. This set includes models I, II, and IV of \cite{Ramond:1993kv} and more general (11 = 13 = 31 = 0) 4-zero
texture models considered for example in GUTs.

\subsubsection{Models with non-zero 13 elements \label{sec:13neq0}}

In this section, we show that the addition of non-zero 13 elements in either $Y^U$ or $Y^D$ allows $\sin 2\b$
and $|V_{ub}/V_{cb}|$ to increase enough to be consistent with the data.  We choose to concentrate on two of the
more constrained examples of these types of textures: models III and V listed in Table \ref{t:5textures}. We will
see that these models are consistent with current data, implying that it is also true for less constrained models
of this type.   
This result also agrees with that found by Roberts et al.~\cite{Roberts:2001zy}.

The important similarities between models III and V are that both have $Y^U_{13} \neq 0$ and $Y^U_{12} = 0$. Each
has 2 non-removable phases, and for both models we place one of these in $Y^U_{13}$. The other we place in
$Y^D_{32}$ ($Y^U_{32}$) for model III (V).  Vanishing $Y^U_{12}$ implies $s^U_{12} = 0$ and the CKM matrix is
\bea V_{CKM} & = &
\left( \ba{ccc} 1 & s^*_{12} & s^*_{13} \\
-s_{12} & 1 & s^*_{23} \\
s^D_{12} s_{23} -s_{13} \ \ \ & -s_{23} - s^{D*}_{12} s_{13} & 1 \ea \right) . \eea Three CKM elements are in one
to one correspondence with $s_{ij}$.  We find  \bea
V_{us} & = & s^*_{12} = \sqrt{\f{m_d}{m_s + m_d}}, \\
V_{ub} & = & s^*_{13} = \sqrt{\f{m_u}{m_t}} e^{-i\phi^U_{13}}, \nn \\
V_{cb} & = & s^*_{23} \nn \eea  where we use equation (\ref{eq:Vus1}) for $|V_{us}|$.\footnote{For model III, the
phase in $Y^D_{32}$ enters in $V_{us}$ through $\tilde{Y}^D_{22}$. However, since $\tilde{Y}^D_{22}$ is dominated
by the real $Y^D_{22}$, the overall phase of $\tilde{Y}^D_{22}$ is small. Finally, since $V_{us} \simeq
Y^D_{12}/\tilde{Y}^D_{22}$, this implies that $V_{us}$ has a negligible phase.}  We thus obtain predictions for
$V_{us}$ and $V_{ub}$ in terms of quark mass ratios. The predicted values, $|V_{us}| = 0.224 \pm 0.008$ and
$|V_{ub}| = 0.0031 \pm 0.0005$, agree quite well with experiment.

\begin{figure}[t]
\begin{center}
\begin{picture}(100,200)
\put(10,105){\makebox(50,50){\includegraphics[height= 3.5
in]{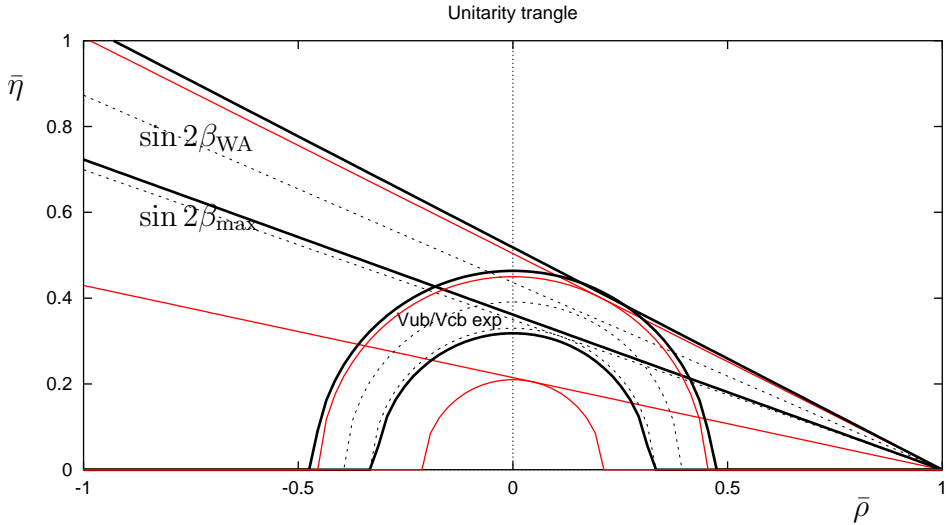}}} \put(-150,160){$\bar\eta$} \put(-100,140){$\sin
2\b_{\rm WA}$} \put(-100,110){$\sin 2\b_{\rm max}$}
\put(170,0){$\bar\rho$}
\end{picture}
\caption{Comparisons between experiment and predictions
of models III and V. Pictured is the predicted $\sin 2\b_{\rm max}$
with red (or grey) lines for the 2$\sigma$ errors and the
experimental 2$\sigma$ lines for $\sin 2\b$ shown in black.
Similarly, 2$\sigma$ arcs are shown in red (or grey) for
the predicted $|V_{ub}/V_{cb}|$, while the experimental 2$\sigma$
arcs are black. There is significant overlap between the
data and predicted values.}
\label{fig:sin2bII}
\end{center}
\end{figure}

$\b$ is given by \bea \b & = & {\rm arg} \left( \f{1}{1 - \f{s_{13}}{s^D_{12} s_{23}}} \right) = \arg \left( 1 -
\rho \ e^{-i \phi^U_{13} }\right) \eea where \bea \rho \equiv \left| \f{V_{ub}}{V_{cb} \ V_{us}} \right|. \eea
The maximum value of $\sin 2\b$ obtained from these models is \bea
\sin \b_{\rm max} & = & \rho = 0.33 \pm 0.06 \\
\sin 2\b_{\rm max} & = & 0.62 \pm 0.10. \nn
\eea
This value agrees with the measured $\sin 2\b$ within 2$\sigma$. There
are no further predictions to be made from models III and V.

Figure \ref{fig:sin2bII} compares the experimental and predicted values
for models III and V. There is clearly a large amount of overlap between
the predictions and experiment, showing that these two models are
in agreement with current measurements. Because III and V represent rather constrained examples
of a broader class of models, those models with non-zero $Y^U_{13}$ or $Y^D_{13}$
that have fewer texture zeros will also be consistent with the data.

\section{Conclusion}

In this paper we have considered hierarchical quark mass textures with symmetric $12 = 21$ and vanishing $11$
elements.  In Sections \ref{sec:13=0} (\ref{sec:13neq0}) 
we addressed textures with $13 = 0$ ($\neq 0$)
elements, respectively.  In the first class of textures, $\sin\beta_{max}$ is given by \bea r = \left|
\f{V_{ub}}{V_{cb}} \right| \sqrt{\f{m_s}{m_d}} \approx  \left| \f{V_{ub}}{V_{cb} \ V_{us}} \right| ,   \eea while
in the second class of textures  $\sin\beta_{max}$ is given by \bea \rho =  \left| \f{V_{ub}}{V_{cb} \ V_{us}}
\right| .   \eea  In both cases, the maximum value of $\sin 2\b$ is determined once $|V_{us}|$, $|V_{cb}|$ and
$|V_{ub}|$ are known. The determination of $|V_{us}|$ and $|V_{cb}|$ is the same for both cases. The difference
comes from the prediction for $|V_{ub}|$. For the textures with $11=13=0$ and $12=21$, we have $|V_{ub}/V_{cb}| =
\sqrt{m_u/m_c}$.  On the other hand, the second class of textures (with $13 \neq 0$) and, in particular, types
III and V 5-zero textures, give $|V_{ub}| = \sqrt{m_u/m_t}$ and $|V_{ub}/V_{cb}| = \sqrt{m_u/m_c}
\f{\sqrt{m_c/m_t}}{|V_{cb}|}$.  It turns out that $\f{\sqrt{m_c/m_t}}{|V_{cb}|} \simeq 1.5$ if we take the
central values of $m_c$, $m_t$ and $|V_{cb}|$.   As a result the hierarchical textures with $13 \neq 0$ are still
consistent with the data.

We have shown that hierarchical texture zero models with $11=13=0$ and $12=21\neq0$ are ruled out at the
3$\sigma$ level by measurements of $\sin 2\b$ and $|V_{ub}/V_{cb}|$. Included in these are the 5-zero models I,
II and IV of \cite{Ramond:1993kv}, and also many GUT models based on SO(10) with 4-zero textures. Allowing
non-zero 13 elements alleviates the conflict, and as examples we have shown that the 5-zero models III and V of
\cite{Ramond:1993kv} predict $|V_{us}|$, $|V_{ub}|$, and $\sin 2\b_{\rm max}$ values which are consistent with
the data. Lastly, we expect that among the observables, $\sin 2\b$ will become a dominant force in constraining
texture zero models in the future.

\section*{Acknowledgment}

We thank A. V. Manohar and A. S. Kronfeld for discussions on the quark mass error estimates.

\appendix

\section*{Appendix}

\section{Loop Rephasing Method}

The purpose of this Appendix is to present a method
to easily count and place the non-removable
phases of the quark mass matrices.
Although Kusenko and Shrock \cite{Kusenko:1994wu} have already given a procedure for this task,
we feel that our method is in practice
easier and faster to use.

For ease of reference, we present our method
in a concise form here. Explanation and definitions
of terms will follow.

\begin{quote}
The loop rephasing method: Form the combined quark mass matrix and find all of its loops.
The number of loops in a set of basis loops is equal to the number
of non-removable phases. The non-removable phases may be
placed on any non-zero elements subject to the constraint that every loop
has at least one corner with a phase.
\end{quote}

A more thorough description follows, including definitions
of the combined quark matrix, loops, corners, and sets of basis loops.
Throughout this description we will be considering as
an example model III from Table \ref{t:5textures}:
\bea
Y^U = \left( \ba{ccc} 0 & 0 & X \\
0 & X & 0 \\
X & 0 & X \ea \right) \;\;
Y^D = \left( \ba{ccc} 0 & X & 0 \\
X & X & X \\
0 & X & X \ea \right)
\eea

We define the combined quark matrix $\overline{Y}$
as the 3 by 6 matrix formed when adjoining
the up and down quark matrices. The 3 rows correspond to the 3 families
of $SU(2)$ doublet fields $Q_i$, and the 6 columns correspond to the
$SU(2)$ singlet fields $(u^c,d^c)_k$.
\bea
Q_i \overline{Y}_{ij}
\left(
\begin{array}{c}
u^c \\
d^c
\end{array}
\right)_j & = &
\left( Q_1, \ Q_2, \ Q_3 \right)
\left(
\begin{array}{cccccc}
X & X & X & X & X & X \\
X & X & X & X & X & X \\
X & X & X & X & X & X \\
\end{array}
\right)
\left(
\begin{array}{c}
u^c_1 \\
u^c_2 \\
u^c_3 \\
d^c_1 \\
d^c_2 \\
d^c_3
\end{array}
\right)
\eea
Although we are concerned specifically with the rephasing
of combined quark matrices with three generations, 
the method presented here
can be applied to any finite complex matrix whose
fields can be rephased independently.

For model III, the combined quark matrix is
\bea
\overline{Y}_{{\rm III}} =
\left(
\ba{cccccc}
0 & 0 & X & 0 & X & 0 \\
0 & X & 0 & X & X & X \\
X & 0 & X & 0 & X & X
\ea
\right)
=
\left(
\ba{cccccc}
0 & 0 & Y^U_{13} & 0 & Y^D_{12} & 0 \\
0 & Y^U_{22} & 0 & Y^D_{21} & Y^D_{22} & Y^D_{23} \\
Y^U_{31} & 0 & Y^U_{33} & 0 & Y^D_{32} & Y^D_{33}
\ea
\right).
\eea

Consider horizontal and vertical lines which connect
non-zero matrix elements. We define a loop to be
a set of such lines which traces a closed path, provided
that no two lines are on top of one another.
We further define that a corner element of a loop
be an element where the loop path makes a $90^\circ$ turn.

In the combined quark matrix $\overline{Y}_{{\rm III}}$, we may form 3
loops:
\begin{center} \begin{picture}(300,65)(5,15)
\Line(24.5,67)(53.5,67)
\Line(53.5,67)(53.5,35)
\Line(53.5,35)(24.5,35)
\Line(24.5,35)(24.5,67)
\Text(31.75,19)[]{${\rm loop}_1$}

\Line(171.5,51)(186,51)
\Line(186,51)(186,35)
\Line(186,35)(171.5,35)
\Line(171.5,35)(171.5,51)
\Text(149.25,19)[]{${\rm loop}_2$}

\Line(259.5,67)(288.5,67)
\Line(288.5,67)(288.5,51)
\Line(288.5,51)(303,51)
\Line(303,51)(303,35)
\Line(303,35)(259.5,35)
\Line(259.5,35)(259.5,67)
\Text(266.75,19)[]{${\rm loop}_3$}

\stepcounter{equation}
\Text(376,51)[r]{(\theequation)}

\Text(150,50)[]{$
\left(
\ba{cccccc}
0 & 0 & X & 0 & X & 0 \\
0 & X & 0 & X & X & X \\
X & 0 & X & 0 & X & X
\ea
\right)
\;\;
\left(
\ba{cccccc}
0 & 0 & X & 0 & X & 0 \\
0 & X & 0 & X & X & X \\
X & 0 & X & 0 & X & X
\ea
\right)
\;\;
\left(
\ba{cccccc}
0 & 0 & X & 0 & X & 0 \\
0 & X & 0 & X & X & X \\
X & 0 & X & 0 & X & X
\ea
\right)
$}
\end{picture} \end{center}

For each loop there is a corresponding rephase-invariant combination
of elements.
At each corner of the loop there is a non-zero matrix
element. Multiply these corner elements together with the
prescription that as the loop is traversed the elements should
be taken as alternating with and without a complex conjugate.
The phase of this product is the phase of the loop.
Our convention is that we start on the left end of the upper
row of the loop and take this element without a complex conjugate.

We now prove that the phase corresponding to a loop is invariant
under rephasing.
Consider a matrix $M$ which contains loops. Consider further one of those
loops, and further still a row (call it the $i$th row) which contains
corners of that loop.
Along this row, the corners of the loop must
be connected pair-wise by horizontal lines. This follows from the
definition of a loop and of loop corners. Let $\theta$ be the
phase associated with the loop:
\bea
\theta = \arg (\ldots M_{ij} M^*_{ik} \ldots M_{il} M^*_{im} \ldots)
\eea
where we have shown only those corner pairs in the $i$th row. Rephase
the $i$th row field by $e^{i \phi}$ so that the new matrix has
the $i$th row elements: $M^\prime_{ij} \equiv e^{i \phi} M_{ij}$ for all $j$.
The phase of the loop in the new matrix will be:
\bea
\theta^\prime & \equiv & \arg (\ldots M^\prime_{ij} M^{\prime*}_{ik} \ldots M^\prime_{il} M^{\prime*}_{im} \ldots ) \\
& = & \arg ( \ldots  \left( e^{i \phi} M_{ij} \right) \left( e^{i \phi} M_{ik} \right)^*
 \ldots \left( e^{i \phi} M_{il} \right) \left( e^{i \phi} M_{im} \right)^* \ldots ) \nn \\
& = &  \arg ( \ldots  M_{ij} M^*_{ik} \ldots M_{il} M^*_{im} \ldots ) \nn \\
& = & \theta. \nn
\eea
These same arguments clearly hold for columns as well. Therefore,
rephasing any row or column cannot change $\theta$ and so
the phase of the loop is a rephase-invariant quantity.

For model III the phases corresponding to the three loops are:
\bea
\theta_1 & = & \arg (Y^U_{13} \ Y^{D*}_{12} \ Y^D_{32} \ Y^{U*}_{33} ) \\
\theta_2 & = & \arg (Y^D_{22} \ Y^{D*}_{23} \ Y^D_{33} \ Y^{D*}_{32} ) \nn \\
\theta_3 & = & \arg (Y^U_{13} \ Y^{D*}_{12} \ Y^D_{22} \ Y^{D*}_{23} \ Y^D_{33} \ Y^{U*}_{33} ). \nn
\eea
Note that $Y^D_{32}$ does not appear in the expression for $\theta_3$ since the element
does not lie on a corner of the loop.

The phases of the loops of a matrix are not necessarily
independent. This can be seen in our example: $\theta_1 + \theta_2 =
\theta_3$. There is a parallel between adding phases
and combining loops. To add a loop, take the
loop in a clockwise direction starting from the left end
of the upper row of the loop. To subtract a loop, take the loop
in the counter-clockwise direction starting from the same place. When
adding or subtracting loops, if two loops
have congruent lines with opposite directions, cancel these
lines out. The resulting loop will correspond to a phase
which is the sum (or difference) of phases of the input loops.

For the specific case of model III, we can see that ${\rm loop}_1
+ {\rm loop}_2 = {\rm loop}_3$:
\begin{center} \begin{picture}(300,45)(5,30)

\Line(78.5,67)(90,67)
\ArrowLine(90,67)(107.5,67)
\ArrowLine(107.5,67)(107.5,51)
\ArrowLine(107.5,51)(107.5,38)
\Line(107.5,38)(107.5,35)
\Line(107.5,35)(96,35)
\ArrowLine(96,35)(78.5,35)
\Line(78.5,35)(78.5,48)
\ArrowLine(78.5,48)(78.5,67)

\ArrowLine(107.5,51)(122,51)
\ArrowLine(122,51)(122,35)
\ArrowLine(122,35)(107.5,35)
\ArrowLine(107.5,35)(107.5,48)
\Line(107.5,48)(107.5,51)

\Line(206,67)(217.5,67)
\ArrowLine(217.6,67)(235,67)
\ArrowLine(235,67)(235,51)
\ArrowLine(235,51)(249.5,51)
\ArrowLine(249.5,51)(249.5,35)
\ArrowLine(249.5,35)(206,35)
\Line(206,35)(206,48)
\ArrowLine(206,48)(206,67)

\stepcounter{equation}
\Text(376,51)[r]{(\theequation)}

\Text(150,50)[]{$
\left(
\ba{cccccc}
0 & 0 & X & 0 & X & 0 \\
0 & X & 0 & X & X & X \\
X & 0 & X & 0 & X & X
\ea
\right)
\rightarrow
\left(
\ba{cccccc}
0 & 0 & X & 0 & X & 0 \\
0 & X & 0 & X & X & X \\
X & 0 & X & 0 & X & X
\ea
\right)
$}
\end{picture} \end{center}

We define a set of basis loops for a matrix as the minimal set of loops
from which all loops of that matrix can be made by addition and subtraction.
For the following, let the number of loops in a basis be equal to $N$.
We will now prove claims relating $N$ to the number and placement of the
non-removable phases.

First, we claim that the number of phases necessary to parametrize the invariant phases of a matrix is equal to
$N$. Our proof: Given a basis of loops for a matrix, all possible loops of that matrix can be made from linear
combinations of these basis loops. Similarly, all invariant phases associated with the loops can be parametrized
by the basis loop phases. As proven later in this Appendix, any loop-free matrix can be made real by rephasing.
Thus all invariant phases are associated with loops, 
and all invariant phases in a matrix are
parametrized by the $N$ phases of the basis loops.

Our second claim is that it is
always possible to place the $N$ non-removable phases into
$N$ different elements of the matrix while choosing 
the rest of the matrix elements to be real.
As proof we give a procedure by which we place these $N$ phases into a configuration
reachable by rephasing a matrix of complex elements. For the following,
we assume $N \geq 3$. The procedure clearly works for $N=1$ or $2$.

1) Choose any loop, then choose
any corner of that loop. Place a phase on that corner and choose all
other corners of that loop to be real.

2) Choose another loop, independent of the first,
which does not contain the corner where we have placed the first phase.{\footnote{It is always possible to choose such a loop because: 
a) it is possible to choose a loop that is independent from
the first (because $N>1$), and b) if this second loop {\em did} contain the
phased corner, we could find a third loop (independent from the first) 
which did not contain that corner by adding/subtracting
the first and second loops.}}
Because this loop
is independent from the first, it must contain new corners not contained in the first loop. Among these
new corners, choose one to contain a phase and choose the rest to be real.

3) Choose another loop, independent of the previous loops, which contains no phased corners. Again,
it is possible to choose such a loop by combining any new, independent loop with
the previous loops in such a way that we achieve a new loop which contains no previously phased corners.
Once we have our desired new loop, among the new corners reached by this loop give one corner a phase and choose the
rest to be real.

4) Repeat step 3) until there have been $N$ loops chosen and $N$ phases placed.

5) Choose as real all elements which are not corners of any loops.

Since they are independent of each other, the chosen set of loops will form a basis.
Further, the set will be such that each loop in the basis has exactly one corner with a phase.
This implies that each basis phase is in one to one correspondence
with an arbitrary matrix element phase. Therefore, by altering these arbitrary matrix phases,
we can independently alter the basis phases.

All invariant loop phases $\theta_i$ can be parametrized by these basis phases $\widehat{\theta}_\a$:
\bea
\theta_i = \sum_{\a=1}^N a_{i \a} \widehat{\theta}_\a.
\eea
The $a_{i \a}$ have values $\pm 1$ or 0 and depend only on the geometry of the matrix loops.
For any given $\theta_i$, we can alter the independent and arbitrary $\widehat{\theta}_\a$ to
give $\theta_i$ any value we choose. This implies our phase choices have
resulted in a matrix in which every loop
contains a corner with a phase. We show later that such a situation can be achieved
by rephasing the fields of the matrix starting from arbitrary complex entries.
Therefore, starting from a matrix of arbitrary complex elements,
we have shown that it is possible to place the $N$ non-removable phases
in $N$ different elements of the matrix, while making all other elements real.

In our example involving model III, any two of the loops may be considered
as a basis set and so there are two non-removable phases. As a parallel, it
is also clear that any two of the $\theta_i$ may combine
to give the third.

Now to determine where the non-removable phases can be placed.
Our claim here is that to have a correct rephasing of the matrix
it is necessary and sufficient
to place phases in the matrix in such a
way that all loops have at least one corner with
a phase. As stated above, the minimal number of phases
necessary to do this is equal to $N$, the number of loops
in a set of basis loops.

Argument for the necessity of our claim:
If after placement of phases there is a loop with no
phases on corner elements, then the invariant phase corresponding
to that loop is zero.
However, starting from a complex matrix with arbitrary
phases there is no way to rephase the fields to
set an invariant phase to a specific value (zero here).
Therefore, for the rephasing to be correct, each
loop must have at least one corner with a phase.

Argument for the sufficiency of our claim:
Let our matrix be A. Place phases in A as described
above so that each loop has at least one corner
with a phase. Form a new matrix B from A by setting
to zero those elements where phases were just placed.
By definition, B must have no loops.
There exists a rephasing procedure, described below,
by which all elements in a loop-free matrix such as B may be made real.
The same rephasing procedure on a matrix of the form A with arbitrary
phases in all elements will lead to the placement of phases
we achieved by our method.
Therefore, the phase placement chosen for A results in a
configuration which can be achieved by rephasing a generally
complex matrix A.

The rephasing procedure for a loop-free matrix is as follows.
First draw all possible horizontal and vertical lines between the non-zero elements
of the matrix. Define the term tree to refer to each set of connected paths.
Choose a tree, and choose a row or column of the matrix which contains an element
of that tree. For simplicity of language, we imagine starting with a row.
For all non-zero elements in the row (all must be connected in this single tree), use the
corresponding column field phases to make these elements real.
Now find all other elements within the columns just used in rephasings. Note that
each of these elements are connected to the previous elements by vertical lines. Because their
column fields have been used already, these elements must be rephased
by using their individual row fields.
This is possible since no two of these elements may be in the same row,
a fact which follows from the loop-free structure of the matrix and is proven later.

Now find all other elements in the rows just used. Note again that
these newly found elements are connected to those just rephased by
horizontal lines.
Rephase these new elements by their column fields.
Continue this procedure, alternating between rows and columns,
until all of the elements within the tree are real.

Now we show that during our procedure, when rephasing in rows (columns), no two elements
to be rephased can be in the same row (column). For simplicity, we take the
case of rephasing in rows. The argument clearly also applies to columns.

Suppose that when rephasing by rows, two of the elements to be rephased lie in the same row.
Because our procedure rephases elements connected by lines to those we have
rephased before, any element we reach in the procedure has a
path back to the original (connected) row.
Our two elements we are considering are connected to each other through two paths: the horizontal
line between them, and the path which goes through the starting
row of the procedure. The existence of two paths connecting two elements
implies the existence of a loop in our supposed loop-free matrix.
The contradiction leads us to conclude that our supposition is false.
Therefore, when rephasing by rows, no two elements to be rephased will
lie in the same row.

Our procedure reaches the whole tree in a finite number of steps because (by assumption) the matrix
is finite which means the tree is finite, and because each step necessarily reaches
more of the tree.
Other trees within the matrix, by definition, have no elements on the
rows or columns used by any other trees. Therefore, the trees of a matrix
may be rephased independently, one after another, by the procedure described
above.

As for our model III example, in section \ref{sec:13neq0} we chose to place
the non-removable phases in $Y^U_{13}$ and $Y^D_{32}$. This places a phase
on at least one corner of each loop. Removal of the elements
at which we have placed phases leads us to
\begin{center} \begin{picture}(300,45)(5,30)

\Line(191,35)(264,35)
\Line(264,35)(264,51)
\Line(264,51)(205.5,51)
\Line(249.5,51)(249.5,67)

\stepcounter{equation}
\Text(376,51)[r]{(\theequation)}

\Text(150,50)[]{$
\left(
\ba{cccccc}
0 & 0 & 0 & 0 & Y^D_{12} & 0 \\
0 & Y^U_{22} & 0 & Y^D_{21} & Y^D_{22} & Y^D_{23} \\
Y^U_{31} & 0 & Y^U_{33} & 0 & 0 & Y^D_{33}
\ea
\right)
=
\left(
\ba{cccccc}
0 & 0 & 0 & 0 & X & 0 \\
0 & X & 0 & X & X & X \\
X & 0 & X & 0 & 0 & X
\ea
\right).
$}
\end{picture} \end{center}
There is one tree. We now
go through our rephasing procedure for which we choose to
begin with the bottom row.
Use columns 1, 3 and 6 to make $Y^U_{31}$, $Y^U_{33}$ and $Y^D_{33}$ real.
The only other non-zero entry in these three columns is $Y^D_{23}$, which we
rephase by row 2. In this row, there are three other non-zero entries: $Y^U_{22}$, $Y^D_{21}$
and $Y^D_{22}$. Rephase these by columns 2, 4 and 5 respectively. There is one other
element, $Y^D_{12}$, in these columns. Rephase it by row 1. There are no other
elements in this row, and we have reached the end of the tree.

When performed on the combined quark matrix $\overline{Y}_{{\rm III}}$ starting
from an arbitrary set of phases on each element,
this same rephasing procedure will lead to the desired
situation of having phases on only $Y^U_{13}$ and $Y^D_{32}$.
Note that this phase choice is not unique.  For example, we could instead have placed phases on $Y^U_{33}$ and $Y^D_{23}$.



\end{document}